\documentclass{article}
\usepackage[T1]{fontenc} 
\usepackage[utf8]{inputenc} 
\usepackage{ismir,cite,url}
\usepackage{graphicx}
\usepackage{color}
\usepackage{booktabs}
\usepackage{subcaption}
\usepackage{multirow,bigdelim}

\usepackage{pifont}

\RequirePackage{amsmath} 
\RequirePackage{amsfonts} 

\usepackage[bookmarks=false,hidelinks]{hyperref}

\newcommand{\iec}{i.\,e.,\,}

\newcommand{\egc}{e.\,g.,\,}

\newcommand{\Sect}[1]{Section~\ref{#1}}

\newcommand{\Fig}[1]{Figure~\ref{#1}}

\newcommand{\Table}[1]{Table~\ref{#1}}

\title{Learning Soft-Attention Models for Tempo-invariant Audio-Sheet Music Retrieval}

\multauthor
{Stefan Balke$^1$ \hspace{0.5cm} Matthias Dorfer$^1$ \hspace{0.5cm} Luis Carvalho$^1$ \hspace{0.5cm} Andreas Arzt$^1$ \hspace{0.5cm} Gerhard Widmer$^{1,2}$}
{$^1$ Institute of Computational Perception, Johannes Kepler University Linz, Austria\\
$^2$ The Austrian Research Institute for Artificial Intelligence (OFAI), Austria\\
{\tt\small stefan.balke@jku.at}
}
\def\authorname{Stefan Balke, Matthias Dorfer, Luis Carvalho, Andreas Arzt, Gerhard Widmer}

\begin{document}

\maketitle
\begin{abstract}
Connecting large libraries of digitized audio recordings to their corresponding
sheet music images has long been a motivation
for researchers to develop new cross-modal retrieval systems.
In recent years, retrieval systems based on embedding space learning with deep neural networks
got a step closer to fulfilling this vision.
However, global and local tempo deviations in the music recordings still require careful
tuning of the amount of temporal context given to the system.
In this paper, we address this problem by introducing an additional soft-attention mechanism on the audio input.
Quantitative and qualitative results on synthesized piano data indicate that this attention
increases the robustness of the retrieval system by focusing on different parts
of the input representation based on the tempo of the audio.
Encouraged by these results, we argue for the potential of attention models as a very general tool for many MIR tasks.
\end{abstract}
%

\section{Introduction}\label{sec:intro}

\begin{figure*}[t]
  \centering
  \includegraphics[width=0.8\textwidth]{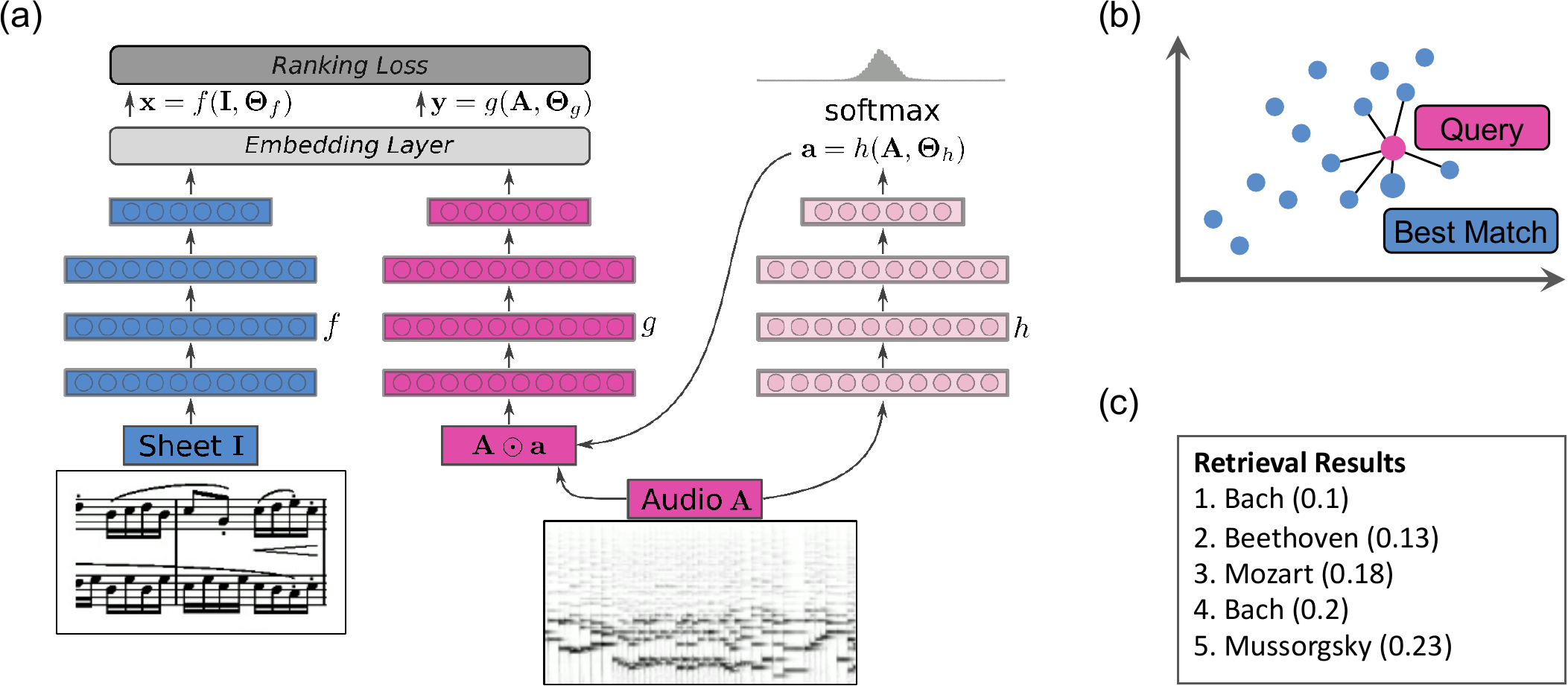}
  \caption{Illustration of the cross-modal retrieval application.
  (a) Audio snippet serves as input query to the embedding network.
  The attention network selects the ``relevant'' region of the audio spectrogram.
  (b) 2-dimensional visualization of the embedding space.
  The nearest neighbors are selected as candidates.
  (c) Ranked list of candidates sorted by their distances to the query embedding.}
\label{fig:teaser}
\end{figure*}


Algorithms for content-based search and retrieval play an important role in many applications that are based on large collections of music data.
In this paper, we re-visit a challenging cross-modal retrieval problem, namely audio--sheet music retrieval: given a short audio excerpt, we are trying to retrieve the corresponding score from a database of sheet music (stored as images).


Traditional methods for audio--sheet retrieval
are based on common mid-level representations that allow for an easy comparison of
time points in the audio and positions in the sheet music,
see for instance
\cite{FremereyCME09_SheetMusicID_ISMIR, IzmirliS12_PrintedMusicAudio_ISMIR,%
KurthMFCC07_AutomatedSynchronization_ISMIR, ArztBW12_SymbolicFingerprint_ISMIR}.
Typical examples are spectral features like pitch class profiles (chroma-features) or
symbolic representations.
Unfortunately, deriving such mid-level representations is an error-prone process,
as this may involve preprocessing steps such as music transcription
\cite{BoeckS12_TranscriptionRecurrentNetwork_ICASSP, ChengMBD16_AttackDecayModel_ISMIR,%
KelzDKBAW16_FramewisePianoTrans_ISMIR, SigtiaBD16_DNNPolyPianoTrans_TASLP,%
HawthorneESRSREOE18_OnsetsAndFramesPianoTrans_ISMIR, KelzBW19_ADSRPianoTranscription_ICASSP},
Optical Music Recognition
\cite{RaphaelW11_OMR_ISMIR,Byrd2015_OMR_JNMR,HajicNPP16_TestbedOMR_ISMIR,%
RebeloFPMGC12_OMRStateArtOpenIssues_IJMIR,WenRZC15_OMRNN_PRL}, and sometimes both.
For a recent overview of different cross-modal retrieval approaches,
see~\cite{MuellerABDW19_MusicRetrieval_IEEE-SPM}.


In~\cite{DorferHAFW18_MSMD_TISMIR}, an alternative approach has been proposed
that circumvents these problematic preprocessing steps by
learning embeddings directly from the multi-modal data
(see \Fig{fig:teaser} for a schematic overview).
Given short snippets of audio and their respective excerpts of sheet music images,
a cross-modal neural network is trained to learn an embedding space in which both
modalities are represented as 32-dimensional vectors (\Fig{fig:teaser}a).
The vectors can then be easily compared in the embedding space by means of
a distance function, \egc~the cosine distance (\Fig{fig:teaser}b).
The retrieval results are then selected by sorting the candidates by the obtained
distances (\Fig{fig:teaser}c).
A conceptually similar retrieval approach for videos was presented
in~\cite{ArandjelovicZ17_LookListenLearn_ICCV}.

In essence, the neural network replaces the step of extracting mid-level features by
directly learning
a transformation from the audio and from the sheet music image data to a common vector space.
A limitation of this approach is that the temporal context (or field of view)
of the data in both modalities is of fixed size.
For audio data, this implies that the actual musical content of the window depends on the tempo
a piece is played in.
If it is played in a fast tempo, the audio excerpt will contain a larger amount of musical
content (\iec~more notes) than if the same piece is played slowly. 
Obviously, this can lead to large discrepancies between what the model has seen during training, and
the data it will see during test time.

A possible solution to this problem is to let the network decide the appropriate temporal context for a given audio query by using a separate attention mechanism
\cite{OlahC16_Attention_Distill,ChanJLV16_ListenAddendSpell_ICASSP,%
BahdanauCB14_NeuralMachineTrans_ICLR,SouthallSH17_DrumTransAttention_ISMIR}.
In a recent workshop paper~\cite{DorferHW18_AudioSheetAttention_ICMLMusic}, the concept of using
a soft-attention mechanism in a cross-modal retrieval scenario was presented.
This allowed the network to deal with a much larger temporal context at the input
than without attention.
%
%
In this paper, we substantially extend this idea with systematic and quantitative experiments
with respect to tempo robustness, as well as giving further details about
the used architecture and the training data.
The remainder of this paper is structured as follows.
In \Sect{sec:asr}, we introduce the retrieval task and
explain the necessary steps to approach it with
end-to-end cross-modal embedding learning.
\Sect{sec:exp} reports on systematic experiments that show
the benefits of using a dedicated attention network on audio spectrograms,
to improve retrieval results compared to the current state of the art.
Furthermore, we present qualitative examples to highlight the intuitive
behavior of the attention network and conclude the paper in \Sect{sec:summary}.

\section{Audio--Sheet Retrieval}\label{sec:asr}
%
We consider a cross-modal retrieval scenario (\Fig{fig:teaser}):
given an audio excerpt as a search query, we wish to retrieve
the corresponding snippet of sheet music of the respective piece.
We approach this retrieval problem
by learning a low-dimensional multimodal embedding space ($32$ dimensions)
for both snippets of sheet music and excerpts of music audio (\Fig{fig:teaser}a).
We desire for each modality a projection into a shared space where
semantically similar items of the two modalities are projected close
together, and dissimilar items far apart (\Fig{fig:teaser}b).
Once the inputs are embedded in such a space, 
cross-modal retrieval is performed
using simple distance measures and nearest-neighbor search.
Finally, the retrieval results are obtained by means of a ranked list
through sorting the distances in an ascending order (\Fig{fig:teaser}c).

The embedding space is trained with convolutional neural networks (CNN).
\Fig{fig:teaser}a sketches the network architecture.
The baseline model (without attention) consists of two convolutional pathways:
one is responsible for embedding the sheet music,
and the other for embedding the audio excerpt.
The key part of the network is the canonically correlated (CCA) embedding layer
\cite{DorferSVKW18_CCALayer_IJMIR}.
This layer forces the two pathways to learn representations that
can be projected into a shared space.
The desired properties of this multimodal embedding space
are enforced by training with pairwise ranking loss
(also known as contrastive loss)~\cite{KirosSZ14_VisualSemanticEmbeddings_arxiv}.
This is the basic structure of the model recently described and
evaluated in~\cite{DorferHAFW18_MSMD_TISMIR}.
This attention-less model serves as a baseline in our experiments,
\iec the input audio (or sheet music) has to be sliced into
excerpts of a given size (\egc 2 s).
However, when processing performed music, the temporal context captured by the fixed-size
input excerpts varies, based on the current tempo of the piece.

\begin{table}[t]
\centering
\scalebox{0.8}{
\begin{tabular}{cc}
\toprule
\textbf{Audio (Spectrogram)} & \textbf{Sheet-Image}\\
$92 \times \{42, 84, 168\}$ & $160 \times 200$\\
\midrule
2x Conv($3$, pad-1)-$24$ - BN	    	& \rdelim\}{11}{120pt}[Attention Network]\\
MaxPooling(2)          				&\\
2x Conv($3$, pad-1)-$48$ - BN			&\\
MaxPooling(2)             				&\\
2x Conv($3$, pad-1)-$96$ - BN			&\\
MaxPooling(2)          				&\\
2x Conv($3$, pad-1)-$96$ - BN			&\\
MaxPooling(2)                    		&\\
Conv($1$, pad-0)-$32$ - Linear          &\\
GlobalAvgPooling + Softmax          &\\
Mult(Spectrogram, Attention)            &\\
\midrule
2x Conv($3$, pad-1)-$24$ - BN	    	&	2x Conv($3$, pad-1)-$24$ - BN \\
MaxPooling(2)          				&	MaxPooling(2) \\
2x Conv($3$, pad-1)-$48$ - BN			&	2x Conv($3$, pad-1)-$48$ - BN \\
MaxPooling(2)             				&	MaxPooling(2) \\
2x Conv($3$, pad-1)-$96$ - BN			&	2x Conv($3$, pad-1)-$96$ - BN \\
MaxPooling(2)          				&	MaxPooling(2) \\
2x Conv($3$, pad-1)-$96$ - BN			&	2x Conv($3$, pad-1)-$96$ - BN \\
MaxPooling(2)                    		&	MaxPooling(2) \\
Conv($1$, pad-0)-$32$ - Linear - BN 	&   Conv($1$, pad-0)-$32$ - Linear - BN  \\
Dense($32$) + Linear  	    			&	Dense($32$) + Linear \\
\midrule
\multicolumn{2}{c}{Embedding Layer + Ranking Loss} \\
\bottomrule
\end{tabular}}
\caption{Overview of the network architecture. The upper part describes the attention network,
the lower part the embedding part.
Conv($3$, pad-1)-$24$: 3$\times$3 convolution, 24 feature maps and zero-padding of 1.
We use ELU activation functions on all layers if not stated otherwise~\cite{ClevertUH15_ELU_ICLR}.}
\label{tab:architecture}
\end{table}

This attention-less model trains and operates on fixed-size
input windows from both modalities.
In other words, the musical content provided to the CNN may contain significantly less or
more information---especially note onsets---than excerpts it has been trained on.
One may of course compensate this with data augmentation, but a more
general solution is to simply let the model decide how much information
is needed from the audio.

For this purpose, we explore a soft-attention mechanism.
First, we substantially increase the audio field of view 
(number of spectrogram frames), up to a factor of four.
Next, we add to the model the attention pathway $h$,
which should learn to restrict the audio input again
by focusing only at those parts that appear relevant for an efficient search query.
As detailed in \Table{tab:architecture}, this attention mechanism
is implemented as a separate CNN.
The output of this CNN is a probability density function $a$ which has the same
number of frames as the audio spectrogram.
Before feeding the spectrogram to the audio embedding network $g$,
we multiply each frame with its attention weight.
This enables the model to cancel out irrelevant parts of the query
and focus on the important information.
In the following experiments, we show that adding this attention network
in combination with the longer temporal context, substantially improves
the results in the considered audio--sheet retrieval task.

\section{Experiments}\label{sec:exp}
%
This section reports on the conducted retrieval experiments.
We start by describing the data used for training and testing the models, and the
data augmentation steps applied during training.
Afterwards, we present the results for the two main experiments, both dealing
with audio--sheet retrieval:
given an audio excerpt, retrieve
the corresponding snippet of sheet music of the respective piece.
Finally, we look at the attention-layer's behavior
for five examples and discuss benefits and limitations of the approach.

\subsection{Data Description and Training}\label{subsec:data}

We use a dataset of synthesized classical piano music,
called \emph{MSMD}~\cite{DorferHAFW18_MSMD_TISMIR}.
In version 1.1, MSMD comprises 467 pieces by 53 composers,
including Bach, Mozart, Beethoven and Chopin,
totalling in over a thousand pages of sheet music and 15+ hours of audio,
with fine-grained cross-modal alignment between note onsets and noteheads.
The main changes from version 1.0 (as used in~\cite{DorferHAFW18_MSMD_TISMIR})
to version 1.1 are that we cleaned the test set from broken pieces and
set all note velocities to a value of $64$.
The scores and audio are both synthesized using the score engraving software
LilyPond\footnote{\url{http://www.lilypond.org}} and
FluidSynth.\footnote{\url{http://www.fluidsynth.org}}

During training, we augment the sheet music by resizing and shifting the image.
For augmenting the audio, we vary the tempo between 95 and 110 \% and
sample from a pool of three different piano soundfonts.
For details of these augmentation steps, we kindly refer the
reader to the explanation given in~\cite{DorferHAFW18_MSMD_TISMIR}.
After training convergence, we refine the used CCA embedding layer on the whole training set.
The reason for this is that during training, the covariance estimates used in the CCA projection
are only based on the number of samples contained in the mini-batch.

For testing, we select 10,000 audio--sheet music pairs from an unseen testset,
where the synthesized audio is rendered with a separate, hold-out piano soundfont.
The rendering procedure for the sheet music remains the same.
The above described augmentation steps are disabled during testing.
In terms of performance, the refinement step of the CCA layer yields an increase in performance of around
0.07 up to 0.15 in terms of mean reciprocal rank (MRR) on the test set
(see \Table{tab:exp1_results} for details).
The used dataset
,\footnote{\url{https://github.com/CPJKU/msmd/tree/v1.1}}
as well as the implementation along with
trained models
,\footnote{\url{https://github.com/CPJKU/audio_sheet_retrieval/tree/ismir-2019}}
are publicly available. 

\subsection{Experiment 1: Attention}\label{subsec:exp_1}

\begin{table}[t!]
 \centering
 \begin{subtable}{\columnwidth}
 \centering
 \scalebox{0.97}{
 \begin{tabular}{lcccccc}
 \toprule
 \textbf{Model} & \bfseries R@1 & \bfseries R@5 & \bfseries R@25 & \bfseries MRR & \bfseries MR\\
 \midrule
 BL1-SC~\cite{DorferHAFW18_MSMD_TISMIR} & 13.67 & 34.62 & 57.44 & 0.24 & 15\\
 \midrule
 BL2-SC & 34.25 & 54.68 & 70.62 & 0.44 & 4\\
 BL2-MC & 36.59 & 60.52 & 77.21 & 0.48 & 3\\
 BL2-LC & 27.66 & 51.79 & 70.34 & 0.39 & 5\\
 \midrule
 BL2-SC + AT & 38.41 & 59.95 & 74.36 & 0.48 & 3\\
 BL2-MC + AT & 46.54 & 68.45 & 81.10 & 0.57 & 2\\
 BL2-LC + AT & 53.16 & 74.42 & 85.35 & 0.63 & 1\\
 \bottomrule
\end{tabular}}%
\caption{Un-refined CCA layer.}
\label{tab:res_unref}
\end{subtable}
\newline
\vspace*{0.4cm}
\newline
\begin{subtable}{\columnwidth}
 \centering
 \scalebox{0.97}{
 \begin{tabular}{lcccccc}
 \toprule
 \textbf{Model} & \bfseries R@1 & \bfseries R@5 & \bfseries R@25 & \bfseries MRR & \bfseries MR\\
 \midrule
 BL1-SC~\cite{DorferHAFW18_MSMD_TISMIR} & 19.12 & 44.16 & 66.63 & 0.31 & 8\\
 \midrule
 BL2-SC & 48.91 & 67.22 & 78.27 & 0.57 & 2\\
 BL2-MC & 47.08 & 68.19 & 80.82 & 0.57 & 2\\
 BL2-LC & 43.46 & 68.38 & 82.84 & 0.55 & 2\\
 \midrule
 BL2-SC + AT & 55.43 & 72.64 & 81.05 & 0.63 & 1\\
 BL2-MC + AT & 58.14 & 76.50 & 84.60 & 0.67 & 1\\
 BL2-LC + AT & 66.71 & 84.43 & 91.19 & 0.75 & 1\\
 \bottomrule
\end{tabular}}%
\caption{Refined CCA layer.}\label{tab:res_ref}
\end{subtable}
 \caption{Overview of the experimental results. (a) Lists the results without and (b) with
 a refinement of the CCA layer.
 All experiments used 10,000 candidates and were conducted on MSMD-v1.1
 (R@k = Recall@k, MRR = Mean Reciprocal Rank, MR = Median Rank).}
\label{tab:exp1_results}
\end{table}

In the first experiment, we investigate the influence of the additional attention network.
We systematically increase the temporal context of the audio representation from
a short context (SC, 42 frames $=$ 2.1 s),
over a medium (MC, 84 frames $=$ 4.2 s),
to a long context (LC, 168 frames $=$ 8.4 s).
The results are summarized in \Table{tab:exp1_results}.
As evaluation metrics we use different Recalls
(R@1, R@5, R@25 $\in [0,100]$, higher is better),
the mean reciprocal rank (MRR $\in [0,1]$, higher is better),
as well as the median rank (MR $\in [1,10000]$, lower is better).
In the following discussion, we focus on the results of the refined CCA layer,
as given in \Table{tab:res_ref}.

As a baseline (BL1-SC), we use the network architecture as described
in~\cite{DorferHAFW18_MSMD_TISMIR}, which uses the short context (SC).
For this model, approximately 20\% of the queries are on rank 1 (R@1 = $19.12$) and in almost
70\% of the audio queries, the relevant sheet image is within the first 25 ranks (R@25 = $66.63$).
As a second baseline (BL2-SC), we slightly adapt the original architecture by exchanging
the global average pooling layer (before the embedding layer)
by a dense layer for each modality (see the non-attention part of \Table{tab:architecture} for details).
With this adaptation, the results improve significantly to R@1 = $48.91$,
R@25 = $78.27$, and a median rank MR = 2, instead of MR = 8 for BL1.

By increasing the temporal context on BL2 to medium sized context (BL2-MC),
mean reciprocal rank (MRR = $0.57$) and median rank (MR = $2$) stay unchanged.
When increasing the temporal context to the long context (BL2-LC),
the model degrades in performance, \egc R@1 drops from $48.91$\% for SC to
$43.46$\% for LC and the MRR from $0.57$ to $0.55$.
Adding the attention network to the audio input (BL2-SC + AT) improves the results
by 7, 5, and 3\% for the recalls, as well as 0.05 for the MRR compared to BL2-SC.
The more context is given to the network, the better the performance metrics get,
\egc R@1 improves from $58.14$ (BL2-MC + AT) to $66.71$ (BL2-LC + AT).
The MRR, improves from $0.63$ (BL2-SC + AT), over $0.67$ (BL2-MC + AT),
up to $0.75$ (BL2-LC + AT).

We derive two main observations from these results.
First, optimizing the network architecture is important; dropping the global average pooling
in favour of a fully-connected dense layer lifted the results to another level.
The reason could be that the fully-connected layer better retains the structure of the input
spectrogram than the average pooling and in addition can be more selective on relevant
input parts, \egc by setting weights to zero.
Second, the attention network enables the network to deal with larger temporal context sizes.
From a signal processing perspective, one would expect that more context
(and thus longer queries) would always help
since it increases the specificity of the query.
However, since we squash this information into a 32-dimensional embedding vector,
it seems that too much information (\egc too many onsets in the audio),
actually harms the retrieval quality when not using attention.
\begin{table}[t!]
\centering
 \scalebox{0.85}{
 \begin{tabular}{lccccc}
 \toprule
 \textbf{Model} & $\rho=0.5$ & $\rho=0.66$ & \bfseries $\rho=1$ & $\rho=1.33$ & $\rho=2$ \\
 \midrule
 BL1-SC~\cite{DorferHAFW18_MSMD_TISMIR} & 0.20 & 0.27 & 0.31 & 0.30 & 0.22\\
 \midrule
 BL2-SC & 0.44 & 0.52 & 0.57 & 0.56 & 0.46\\
 BL2-MC & 0.46 & 0.53 & 0.57 & 0.55 & 0.43\\
 BL2-LC & 0.44 & 0.50 & 0.55 & 0.51 & 0.35\\
 \midrule
 BL2-SC + AT & 0.55 & 0.63 & 0.63 & 0.64 & 0.56\\
 BL2-MC + AT & 0.54 & 0.61 & 0.67 & 0.67 & 0.62\\
 BL2-LC + AT & 0.64 & 0.69 & 0.75 & 0.73 & 0.64\\
 \bottomrule
\end{tabular}}
 \caption{Mean Reciprocal Rank (MRR) for different models and different tempo ratios
 $\rho \in \{0.5, 0.66, 1, 1.33, 2\}$.
 For example, $\rho=0.5$ stands for half of the original tempo
 and $\rho=2$ for doubling the tempo.
 The listed models correspond to the refined CCA models as listed in \Table{tab:res_ref}
 with the same test set size of 10,000 candidates.}
\label{tab:results_ti}
\end{table}

\begin{figure}[t]
    \centering
    \includegraphics[width=1\columnwidth]{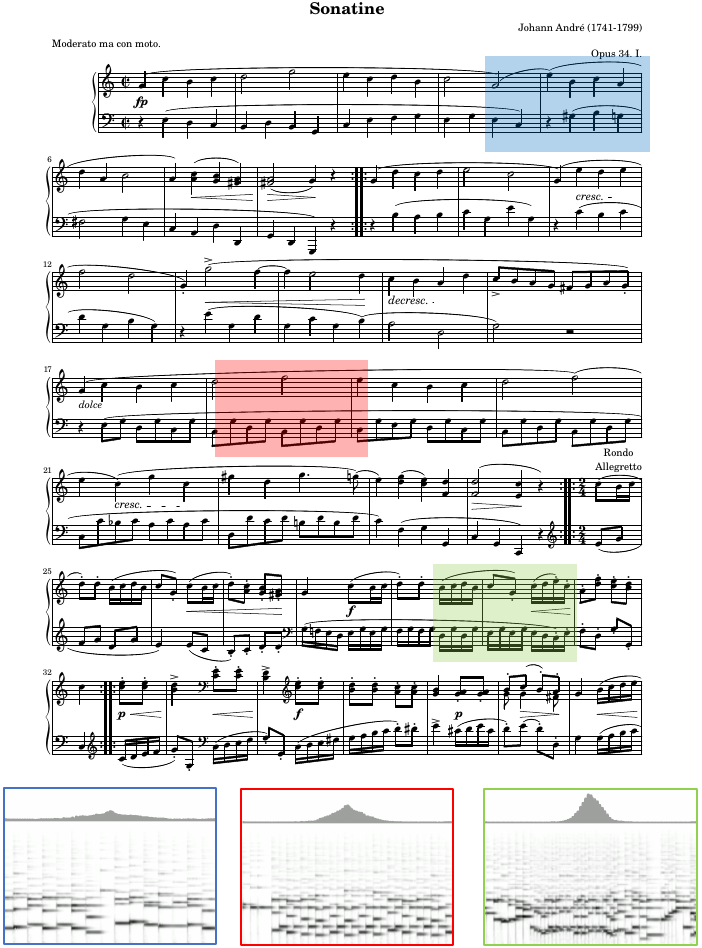}
    \caption{Example of tempo variations within a piece by Johann André - Sonatine (Op. 34, I.).
    The three boxes below the sheet music show the attention output, and the corresponding
    audio spectrogram for the respective excerpt.}
    \label{fig:attention_examples_local}
\end{figure}

\subsection{Experiment 2: Tempo Robustness}\label{subsec:exp_2}
\begin{figure*}[t]
    \centering
    \includegraphics[width=0.9\textwidth]{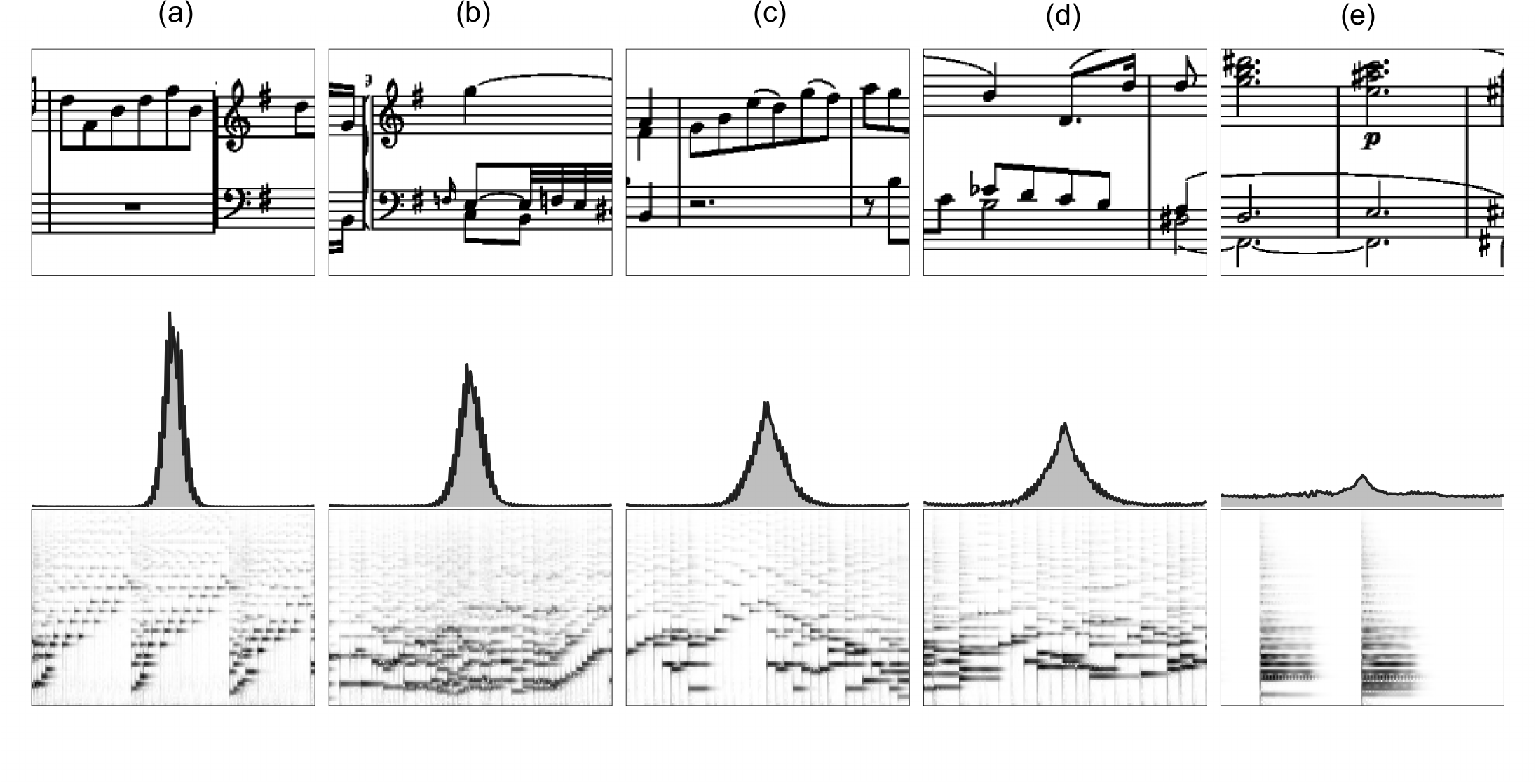}
    \caption{Illustration of the sheet music, input attention (normalized), and spectrogram
    for five examples from the MSMD test set.
    (a) L. v. Beethoven - Piano Sonata (Op. 79, 1st Mvt.),
    (b) J. S. Bach - Goldberg Variations: Variatio 12 (BWV 988),
    (c) J. S. Bach - French Suite VI: Menuet (BWV 817),
    (d) R. Schumann - Album for the Youth: Untitled (Op. 68, Nr. 26), and
    (e) M. Mussorgsky - Pictures at an Exhibition VIII: Catacombae.}
    \label{fig:attention_examples_global}
\end{figure*}

In a second experiment, we explicitly test the system's robustness to global tempo changes.
For this purpose, we re-rendered the MSMD test dataset with various tempo ratios
$\rho \in \{0.5, 0.66, 1, 1.33, 2\}$, where $\rho=0.5$ stands for halving
and $\rho=2$ for doubling the original tempo.
The results are given in \Table{tab:results_ti}. For the sake of brevity,
we only show the mean reciprocal rank (MRR) for the refined CCA models.

In general, we observe a similar trend as in the first experiment.
While the original baseline approach (BL1-SC) performs rather poorly
(0.20 to 0.31), exchanging the global average pooling layer (BL2-SC) helps to improve
the performance (0.44 to 0.57).
The best attention model (BL2-LC + AT) yields values ranging from 0.64 to 0.75.
In this experiment, we would have expected that the improvement holds true for all testsets
rendered at different global tempi.
However, the numbers tell a slightly different story.
To understand this, we had to go back to the generation of the MSMD dataset.
Recall that the dataset is generated from LilyPond files obtained from the
Mutopia Project.\footnote{\url{https://www.mutopiaproject.org}}
These files contain tempo specifications which are retained in the synthesis pipeline.
The specified tempi vary in a range between 30 and 182 bpm
(mean tempo = 106 bpm, std. dev. = 25 bpm).
This distribution of tempi implies that all experiments we perform on the original MSMD testset
($\rho=1$) already test for a variety of tempi.
Synthesizing this dataset with different tempo
factors---as done in our second experiment---shifts and stretches this distribution.
In the edge cases ($\rho=0.5$ and $\rho=2$), this leads to unrealistic tempi,
\egc 15 bpm or 364 bpm, producing input audio windows with absurdly low
or high onset density.

In summary, all of the tested models have in common that they work best
for the original testset tempo ($\rho=1$) and
have similar relative performance drops for the tempo variations.
However, the attention model is able to keep the retrieval results at a much higher level
than all the other models.
In the following section, we take a closer look at some examples from the MSMD test set
to get a better intuition of the soft-attention mechanism and how it reacts to
tempo deviations.

\subsection{Examples}\label{subsec:examples}

In the experiments above, we have seen that the attention models improve the retrieval results
by a considerable margin.
Another positive aspect of the soft-attention mechanism is that its behavior is directly
interpretable---a rare case in the deep learning landscape.
For all the presented examples, we provide further videos with audio on an accompanying website,
along with detailed instructions for reproduction.\footnote{\url{http://www.cp.jku.at/resources/2019_ASR-TempoInv_ISMIR}}

\Fig{fig:attention_examples_local} shows the sheet music for Johann André's Sonatine, Op.34.
The piece starts with a calm part with mainly legato quarter notes at 103 bpm.
The first (blue) box shows the corresponding attention output
and the audio spectrogram for measure five.
The attention output is relatively flat with a small peak in the middle of the audio excerpt.
In bar 17, the note density changes, with eighth notes in the left hand entering the scene.
As shown by the second (red) box, the attention layer reacts to this by building
up a more distinct focus around the center, placing less weight on the outer parts of the spectrogram.
In the third part beginning with measure 24, the ``tempo''
(more precisely: the perceived ``speed'', in terms of events per time unit)
essentially doubles, which is reflected in
a high peak and narrow distribution in the attention output (green right box).
From these three examples, it also becomes clear that
tempo and note density are essentially two sides of the same coin for the
attention network.

In \Fig{fig:attention_examples_global}, we show examples from the testset with different global performance tempi, along with the sheet music excerpt, the attention weights, and the spectrogram.
The fastest piece, with around 250 bpm, is shown in \Fig{fig:attention_examples_global}a.
The corresponding attention output is very focused on the middle part of the audio spectrogram,
trying to concentrate the attention to the notes that actually appear in the sheet music snippet.
The second example (b) is rather slow with 95 bpm.
However, through the use of 16th and 32th notes, the main melody gets a double-time feel, thus the actual tempo is perceived at around 190 bpm.
In \Fig{fig:attention_examples_global}c, the tempo is at 115 bpm.
The attention output starts to widen up, allowing more temporal context to reach the
actual embedding network.
This trend goes in \Fig{fig:attention_examples_global}d when the tempo 78 bpm.
\Fig{fig:attention_examples_global}e shows an extreme example where the piece mainly consists
of chords with long, dotted half notes.
Here, the attention has to use the complete temporal context of the audio spectrogram
to match the score information with the audio.

The examples demonstrate that depending on the spectrogram content, the model indeed attends
to whatever it believes is a representative counterpart of the target sheet music snippet.
Since the fixed-size sheet snippets contain roughly similar 
numbers of notes, as the density of note heads on the printed page tends to be independent of the tempo of the piece,
attention is sharply peaked when the density of onsets in the audio is high,
and conversely is distributed more evenly when there are fewer notes in the audio excerpt.

\begin{figure}[t]
    \centering
    \includegraphics[width=0.95\columnwidth]{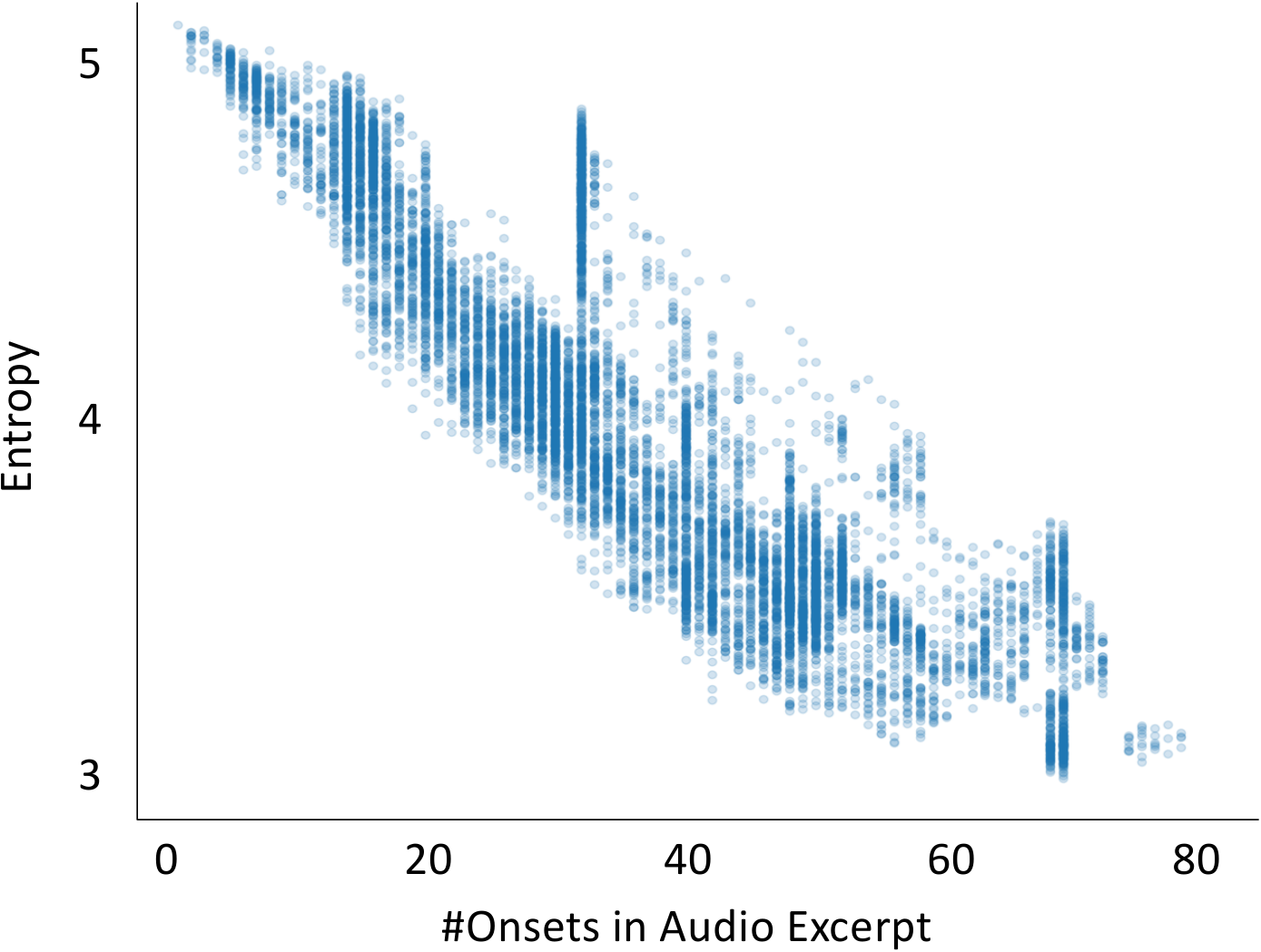}
    \caption{Entropy of the input attention vs. the number of onsets in the
    respective audio frame.}
    \label{fig:attention_entropy}
\end{figure}

\section{Summary}\label{sec:summary}

In this paper, we have described a soft-attention mechanism that helps to
overcome the fixed window sizes uses in Convolutional Neural Networks.
In our end-to-end audio--sheet music retrieval application, the results
improved substantially compared to the state of the art.
By looking at a number of examples from the retrieval results,
the soft-attention mechanism showed an intuitive and interpretable
behavior.

This appealing and intuitive behavior is summarized in \Fig{fig:attention_entropy}, which 
shows the entropy of the attention distribution in relation to the number of onsets contained in the audio excerpt, for all 10,000 test samples.
The entropy is a measure of flatness of the attention output: a flat function gives high entropy, a very narrow function low entropy.
The downward trend in the figure confirms our observations from above:
the more onsets in the respective audio spectrogram,
the narrower the attention distribution.

Given the improved retrieval performance and the intuitive behavior
of the attention model, we think this is a promising line of research
for reducing the sensitivity of cross-modal music retrieval models
to the audio input window size.
To this end, our experiments were conducted on synthesized piano recordings.
However, results in~\cite{DorferHAFW18_MSMD_TISMIR} indicate that
the embedding models trained on this data generalize to real scores and performances.
A possible next step would be to investigate whether the attention mechanism
reacts to local tempo changes as occuring frequently in real performances
(\egc ritardandi and accelerandi).
Furthermore, it would be interesting to leave the piano music domain and
extend the model to cope with differences in timbre,
(\egc orchestral music)
as done in~\cite{BalkeALM16_BarlowRetrieval_ICASSP} for
the challenging Barlow--Morgenstern scenario~\cite{BarlowM75_MusicalThemes_BOOK}.

\vspace{0.5cm}
\noindent\textbf{Acknowledgements}\\
%
This project has received funding from the European Research Council (ERC)
under the European Union's Horizon 2020 research and innovation programme
(grant agreement number 670035, project CON ESPRESSIONE,
and the Marie Skłodowsa-Curie grant agreement number 765068, MIP-Frontiers).
%

\end{document}